\pdfoutput=1 

\documentclass[11pt]{llncs}

\usepackage{amsmath,amssymb}
\usepackage{mathrsfs} 
\usepackage{units}

\newcommand{\FigScale}{0.95}

\usepackage{graphicx} 
\usepackage[tight,footnotesize]{subfigure} 
\usepackage{array} 

\usepackage{url}

\usepackage[noadjust]{cite} 

\def\hyph{-\penalty0\hskip0pt\relax}

\DeclareMathOperator{\oD}{D}                                            
\DeclareMathOperator{\oE}{E}                                            
\DeclareMathOperator{\oH}{H}                                            
\DeclareMathOperator{\oP}{P}                                            
\newcommand{\cR}{\mathcal{R}}                                           

\begin{document}


\title{An Information-Theoretic Privacy Criterion for\\Query Forgery in Information Retrieval}
\author{David Rebollo-Monedero, Javier Parra-Arnau, Jordi Forn\'e}
\institute{Department of Telematics Engineering, Technical University of Catalonia (UPC),
E-08034 Barcelona, Spain \\ \email{\{david.rebollo,javier.parra,jforne\}@entel.upc.edu}%
\thanks{This work was partly supported by the Spanish Government through projects
CONSOLIDER INGENIO 2010 CSD2007-00004 ``ARES",
TEC2010-20572-C02-02 ``CONSEQUENCE",
and by the Government of Catalonia under grant 2009 SGR 1362.}}

\maketitle

\begin{abstract}
In previous work, we presented a novel information\hyph theoretic privacy criterion for query forgery
in the domain of information retrieval. Our criterion measured privacy risk as a divergence between the user's and the
population's query distribution, and contemplated the entropy of the user's distribution as a particular case.
In this work, we make a twofold contribution.
First, we thoroughly interpret and justify the privacy metric proposed in our previous work,
elaborating on the intimate connection between the celebrated method of entropy maximization
and the use of entropies and divergences as measures of privacy.
Secondly, we attempt to bridge the gap between the privacy and the information\hyph theoretic
communities by substantially adapting some technicalities of our original work to reach a wider audience,
not intimately familiar with information theory and the method of types.
\end{abstract}

\setcounter{footnote}{0} 
\renewcommand{\thefootnote}{(\alph{footnote})}  

\section{Introduction}
\label{sec:Intro}

\noindent
During the last two decades, the Internet has gradually become a part of everyday life. One of the most frequent activities
when users browse the Web is submitting a query to a search engine. Search engines allow users to retrieve information on a great
variety of categories, such as hobbies, sports, business or health. However, most of them are unaware of the privacy risks
they are exposed to~\cite{Fallows05PIALP}.

As a concrete example, from November to December of 2008,
61\% of adults in the U.S. looked for online information about a particular disease, a specific
treatment, an alternative medicine, and other related topics~\cite{Fox09PIALP}.
Such queries could disclose sensitive information
and be used to profile users about potential diseases.
In the wrong hands, such private information could be the cause of discriminatory hiring, or could seriously damage someone's reputation.

The fact is that the literature of information retrieval abounds with examples of user privacy threats.
Those include the risk of user profiling not only by an Internet search engine,
but also by location\hyph based service (LBS) providers, or even 
corporate profiling by patent and stock market database providers.
In this context, query forgery, which consists in accompanying genuine
with forged queries, appears as an approach, among many others, to preserve user privacy to a certain extent,
if one is willing to pay the cost of traffic and processing overhead.


In our previous work~\cite{Rebollo10IT}, we presented a novel information\hyph theoretic privacy criterion for query forgery
in the domain of information retrieval.
Our criterion measured privacy risk as a divergence between the user's and the
population's query distribution, and contemplated the entropy of the user's distribution as a particular case.
In this work, we make a twofold contribution.
First, we thoroughly interpret and justify the privacy metric proposed in our previous work,
elaborating on the intimate connection between the celebrated method of entropy maximization
and the use of entropies and divergences as measures of privacy.
Secondly, we attempt to bridge the gap between the privacy and the information\hyph theoretic
communities by substantially adapting some technicalities of our original work to reach a wider audience,
not intimately familiar with information theory and the method of types.


Sec.~\ref{sec:StateOfTheArt} reviews the most relevant approaches in private information retrieval and privacy criteria.
Sec.~\ref{sec:Background:StatisticsAndInformationTheory} examines some fundamental concepts related to information theory
which will help to better understand the essence of this work.
Inspired by the maximum entropy method, we put forth
an information\hyph theoretic criterion to measure the privacy of user profiles in Sec.~\ref{sec:PrivacyCriterion}.
Sec.~\ref{sec:QueryForgery} applies this criterion to the optimization of the trade\hyph off between privacy
and redundancy for query forgery in private information retrieval.
Conclusions are drawn in Sec.~\ref{sec:Conclusion}.

\section{State of the Art in Private Information Retrieval}
\label{sec:StateOfTheArt}

\noindent
Throughout this paper, we shall use the term private information retrieval (PIR) in its widest sense, meaning that we shall
not restrict ourselves to the cryptographically\hyph based techniques normally connected to that acronym.
In other words, we shall refer to the more
generic scenario in which users send general\hyph purpose queries to an information service provider,
say Googling  ``highest\hyph grossing film science fiction?".
Next, we shall introduce the most relevant
contributions to PIR with regard to query forgery and privacy criteria.

\subsection{Private Information Retrieval}
\label{sec:StateOfTheArt:PIR}
\noindent
A variety of solutions have been proposed in information retrieval. Some of them are based on a trusted third party (TTP) acting
as an intermediary between users and the information service provider~\cite{Mokbel06VLDB}.
Although this approach guarantees user privacy thanks to the fact that their identity is unknown to the service provider,
in the end, user trust is just shifted from one entity to another.

Some proposals not relying on TTPs make use of perturbation techniques.
In the particular case of LBS, users may perturb their location information
when querying a service provider~\cite{Duckham01CEUS}.
This provides users with a certain level of privacy in terms of location,
but clearly not in terms of query contents and activity.
Further, this technique poses a trade\hyph off between privacy and data utility: the higher the perturbation of the location,
the higher the user's privacy, but the lower the accuracy of the service provider's responses.
Other TTP\hyph free techniques rely on user collaboration.
In~\cite{Rebollo09IADIS,Rebollo09COMCOM}, a protocol based on query permutation in a trellis of users is proposed,
which comes in handy when neither the service provider nor other cooperating users can be completely trusted.

Alternatively, cryptographic methods for PIR enable a user to privately retrieve the contents from a database indexed by a memory
address sent by the user, making it unfeasible for the database provider to ascertain which
entries were retrieved~\cite{Ostrovsky07PKC,Ghinita08MD}.
Unfortunately, these methods require the provider's cooperation in the privacy
protocol, are limited to a certain extent to query\hyph response functions in the form of a finite lookup table of precomputed answers,
and are burdened with a significant computational overhead.

Query forgery, the focus of our discussion, stands as yet another alternative to the previous methods.
The idea behind this technique is simply to submit original queries along with false queries.
Despite its plainness, this approach can protect user privacy to a certain extent, at the cost of traffic and processing overhead,
but without the need to trust the information provider or the network operator.
Building upon this principle, several PIR protocols, mainly heuristic, have been put forth.
In~\cite{Elovici02WPES, Shapira05IST}, a solution is presented, aimed to preserve the privacy of a group of users sharing
an access point to the Web while surfing the Internet.
The authors propose the generation of fake accesses to a Web page to hinder eavesdroppers in their efforts to profile the group.
Privacy is measured as the similarity between the actual profile of a group of users and
that observed by privacy attackers~\cite{Elovici02WPES}.
Specifically, the authors use the cosine measure, common in information retrieval~\cite{Frakes92PRE},
to capture the similarity between the group's genuine distribution and the apparent one.
Based on this model, some experiments are conducted to study the impact of the construction of
user profiles on the performance~\cite{Kuflik03LNCS}.
In line with this, simple, heuristic implementations in the form of add\hyph ons for popular browsers
have recently appeared~\cite{Howe06B,Toubiana07S}.

Query forgery is also present as a component of other privacy protocols, such as the private
location\hyph based information retrieval protocol via user collaboration in~\cite{Rebollo09COMCOM,Rebollo09IADIS}.
In~\cite{Kido05ICDE}, the authors propose submitting true and false position data when querying an LBS provider,
maintaining certain temporal consistency, rather than doing so completely randomly.


In addition to legal implications, there are a number of technical considerations regarding
bogus traffic generation for privacy~\cite{Soghoian07ISJLP}, as attackers may analyze not only contents but also activity,
timing, routing or any transmission protocol parameters, jointly across several queries or even across diverse information services.
Furthermore, automated query generation is naturally bound to be frown upon by network and information providers,
thus any practical framework must take traffic overhead into account.

\subsection{Privacy Criteria}
\label{sec:StateOfTheArt:Criteria}
\noindent
In this section we give a broad overview of privacy criteria originally intended for statistical disclosure control (SDC),
but in fact applicable to query logs in PIR, the motivating application of our work.
In database privacy, a \emph{microdata set} is defined as a database table whose records carry information
concerning individual respondents.
Specifically, this set contains key attributes, that is, attributes that, in combination, may be linked
with external information to reidentify the respondents to whom the records in the microdata set refer.
Examples include job, address, age and gender, height and weight.
In addition, the data set contains confidential attributes with sensitive information on the respondent,
such as health, salary and religion.

A common approach in SDC is microaggregation, which consists in clustering the data set into groups of records with similar tuples of
key attributes values, and replacing these tuples in every record within each group by a representative group tuple.
One of the most popular privacy criteria in database anonymization is $k$\hyph anonymity~\cite{Samarati98SRI},
which can be achieved through the aforementioned microaggregation procedure.
This criterion requires that each combination of key attribute values be shared by at least $k$ records in the microdata set.
However, the problem of $k$\hyph anonymity,
and of enhancements~\cite{Sun08TDP,Truta06PDM,Machanavajjhala06ICDE,JianMin08ISIP} such as $l$\hyph diversity,
is their vulnerability against skewness and similarity attacks~\cite{Domingo08PSAI}.
In order to overcome these deficiencies, yet another privacy criterion was considered in~\cite{Li07ICDE}:
a dataset is said to satisfy $t$\hyph closeness if for each group of records sharing a combination of key attributes,
a certain measure of divergence between the within\hyph group distribution of confidential attributes and
the distribution of those attributes for the entire dataset does not exceed a threshold~$t$.
An average\hyph case version of the worst\hyph case $t$\hyph closeness criterion,
using the Kullback\hyph Leibler divergence as a measure of discrepancy, turns out to be equivalent to a mutual information,
and lend itself to a generalization of Shannon's rate\hyph distortion problem~\cite{Rebollo08PSD,Rebollo10KDE}.

A simpler information\hyph theoretic privacy criterion, not directly evolved from $k$\hyph anonymity,
consists in measuring the degree of anonymity observable by an attacker as
the entropy of the probability distribution of possible senders of a given message~\cite{Diaz02PET,Diaz05PhD}.
A generalization and justification of such criterion, along with its applicability to PIR, are provided in~\cite{Rebollo10IT,Parra10TB}.

\section{Statistical and Information\hyph Theoretic Preliminaries}
\label{sec:Background:StatisticsAndInformationTheory}

\noindent
This section establishes notational aspects,
and, in order to make our presentation suited to a wider audience,
recalls key information\hyph theoretic concepts assumed to be known in the remainder of the paper.
The measurable space in which a \emph{random variable} (r.v.) takes on values will be called an \emph{alphabet},
which, with a mild loss of generality, we shall always assume to be finite.
We shall follow the convention of using uppercase letters for r.v.'s, and lowercase letters for particular values they take on.
The \emph{probability mass function} (PMFs) $p$ of an r.v.~$X$ is essentially a \emph{relative histogram} across
the possible values determined by its alphabet.
Informally, we shall occasionally refer to the function $p$ by its value $p(x)$.
The \emph{expectation} of an r.v.\ $X$ will be written as $\oE X$, concisely denoting $\sum_x x\,p(x)$,
where the sum is taken across all values of $x$ in its alphabet.

We adopt the same notation for information\hyph theoretic quantities used in~\cite{Cover06B}.
Concordantly, the symbol~$\oH$ will denote entropy and~$\oD$ relative entropy or  Kullback\hyph Leibler (KL) divergence.
We briefly recall those concepts for the reader not intimately familiar with information theory.
All logarithms are taken to base~2.
The \emph{entropy} $\oH(p)$ of a discrete r.v.\ $X$ with probability distribution~$p$ is a measure of its uncertainty,
defined as
$$\oH(X)=-\oE\,\log p(X) = -\sum_x p(x)\log p(x).$$
Given two probability distributions $p(x)$ and $q(x)$ over the same alphabet, the \emph{KL divergence}
or \emph{relative entropy} $\oD(p\,\|\,q)$ is defined as
$$\oD(p\,\|\,q)=\oE_p\,\log \frac{p(X)}{q(X)} = \sum_x p(x) \log \frac{p(x)}{q(x)}.$$
The KL divergence is often referred to as \emph{relative entropy}, as it may be regarded as a
generalization of entropy of a distribution, relative to another.
Conversely, entropy is a special case of KL divergence,
as for a uniform distribution $u$ on a finite alphabet of cardinality~$n$,
\begin{equation}\label{eqn:Background:EntropyAsDivergenceWRTUniform}
\oD(p\,\|u)=\log n - \oH(p).
\end{equation}

Although the KL divergence is not a distance in the mathematical sense of the term,
because it is neither symmetric nor satisfies the triangle inequality,
it does provide a measure of discrepancy between distributions,
in the sense that $\oD(p\,\|\,q)\geq0$, with equality if, and only if, $p=q$.
On account of this fact, relation~\eqref{eqn:Background:EntropyAsDivergenceWRTUniform} between entropy and KL divergence
implies that $\oH(p)\leqslant \log n$, with equality if, and only if, $p=u$.
Simply put, \emph{entropy maximization} is a special case of \emph{divergence minimization},
attained when the distribution taken as optimization variable is identical to the \emph{reference distribution},
or as ``close" as possible, should the optimization problem appear
accompanied with \emph{constraints} on the desired space of candidate distributions.


\section{Entropy and Divergence as Measures of Privacy}
\label{sec:PrivacyCriterion}

\noindent
In this paper we shall interpret entropy and KL divergence as privacy criteria.
For that purpose, we shall adopt the perspective of
Jaynes' celebrated \emph{rationale on entropy maximization methods}~\cite{Jaynes82P},
which builds upon the \emph{method of types}~\cite[\S 11]{Cover06B}, a powerful technique in large deviation theory
whose fundamental results we proceed to review.

The first part of this section will tackle an important question.
Suppose we are faced with a problem, formulated in terms of a model, in which a probability distribution plays a major role.
In the event this distribution is unknown, we wish to assume a feasible candidate.
What is the most likely probability distribution?
In other words, what is the ``probability of a probability" distribution?
We shall see that a widespread answer to this question
relies on choosing the distribution \emph{maximizing the Shannon entropy}, or, if a reference distribution is available,
the distribution \emph{minimizing the KL divergence} with respect to it,
commonly subject to feasibility constraints determined by the specific application at hand.

Our review of the maximum entropy method is crucial because
it is unfortunately not always known in the privacy community,
and because the rest of this paper constitutes a sophisticated illustration of its application,
in the context of the protection of the privacy of user profiles.
As we shall see in the second part of this section, the key idea is
to model a user profile as a histogram of relative frequencies across categories of interest,
regard it as a probability distribution, apply the maximum entropy method to measure
the likelihood of a user profile either as its entropy or as its divergence with respect to the population's average profile,
and finally take that likelihood as a measure of anonymity.

\subsection{Rationale behind the Maximum Entropy Method}
\label{sec:PrivacyCriterion:MaxEnt}

\noindent
A wide variety of models across diverse fields have been explained on the basis of the intriguing principle of entropy maximization.
A classical example in physics is the Maxwell\hyph Boltzmann probability distribution $p(v)$ of particle velocities $V$
in a gas~\cite{Brillouin62B,Jaynes82B} of known temperature.
It turns out that $p(v)$ is precisely the probability distribution maximizing the entropy,
subject to a constraint on the temperature, equivalent
to a constraint on the average kinetic energy, in turn equivalent to a constraint on $\oE V^2$.
Another well\hyph known example, in the field of electrical engineering, of the application of the maximum entropy method,
is Burg's spectral estimation method~\cite{Burg75PhD}.
In this method, the power spectral density of a signal is regarded as a probability distribution of power across frequency,
only partly known.
Burg suggested filling in the unknown portion of the power spectral density by choosing that maximizing the entropy,
constrained on the partial knowledge available.
More concretely, in discrete case, when the constraints consist in a given range of the crosscorrelation function,
up to a time shift~$k$, the solution turns out to be a $k$\textsuperscript{th} order Gauss\hyph Markov process~\cite{Cover06B}.
A third and more recent example, this time in the field of natural language processing, is the use of log\hyph linear models,
which arise as the solution to constrained maximum entropy problems~\cite{Berger96CL} in computational linguistics.

Having motivated the maximum entropy method,
we are ready to proceed to describe Jaynes' attempt to justify, or at least interpret it, by reviewing
the method of types of large deviation theory, a beautiful area lying at the intersection of statistics and information theory.
Let $X_1,\dots,X_k$ be a sequence of $k$ i.i.d.\ drawings of an r.v.\
uniformly distributed in the alphabet $\{1,\dots,n\}$.
Let $k_i$ be the number of times symbol $i=1,\dots,n$ appears in a sequence of outcomes $x_1,\dots,x_k$, thus $k=\sum_i k_i$.
The \emph{type}~$t$ of a sequence of outcomes is the relative proportion of occurrences of each symbol,
that is, the \emph{empirical distribution} $t=\left(\frac{k_1}{k},\dots,\frac{k_n}{k}\right)$, not necessarily uniform.
In other words, consider tossing an $n$\hyph sided fair dice $k$ times, and seeing exactly $k_i$ times face~$i$.
In~\cite{Jaynes82P}, Jaynes points out that
$$\oH(t)=\oH\left(\frac{k_1}{k},\dots,\frac{k_n}{k}\right)
    \simeq \frac{1}{k}\, \log \frac{k!}{k_1! \cdots k_n!} \quad\textnormal{for } k\gg1.$$
Loosely speaking, for large~$k$, the size of a \emph{type class}, that is,
the number of possible outcomes for a given type $t$ (permutations with repeated elements), is approximately $2^{k\oH(t)}$ in the exponent.
The fundamental rationale in~\cite{Jaynes82P} for selecting the type $t$ with maximum entropy $\oH(t)$
lies in the approximate equivalence between entropy maximization and
the maximization of the number of possible outcomes corresponding to a type.
In a way, this justifies the infamous \emph{principle of insufficient reason}, according to which, one
may expect an approximately equal relative frequency $k_i/k=1/n$ for each symbol~$i$, as
the uniform distribution maximizes the entropy.
The principle of entropy maximization is extended to include constraints also in~\cite{Jaynes82P}.

Obviously, since all possible permutations count equally, the argument only works for uniformly distributed drawings,
which is somewhat circular.
A more general argument~\cite[\S 11]{Cover06B}, albeit entirely analogous,
departs from a prior knowledge of an arbitrary PMF $\bar{t}$, not necessarily uniform, of such samples $X_1,\dots,X_k$.
Because the empirical distribution or type $T$ of an i.i.d.\ drawing is itself an r.v.,
we may define its PMF $p(t)=\oP\{T=t\}$; formally, the PMF of a random PMF.
Using indicator r.v.'s, it is straightforward to confirm the intuition that $\oE T = \bar{t}$.
The general argument in question leads to approximating the probability $p(t)$ of a type class, a fractional measure of its size,
in terms of its relative entropy, specifically $2^{-k\oD(t\,\|\,\bar{t})}$ in the exponent, i.e.,
$$\oD(t\,\|\,\bar{t}) \simeq -\frac{1}{k}\,\log p(t)  \quad\textnormal{for } k\gg1,$$
which encompasses the special case of entropy, by virtue of~\eqref{eqn:Background:EntropyAsDivergenceWRTUniform}.
Roughly speaking, the likelihood of the empirical distribution $t$ exponentially decreases with its KL divergence with respect to the
average, reference distribution $\bar{t}$.

In conclusion, the most likely PMF $t$ is that minimizing its divergence with respect to the reference distribution~$\bar{t}$.
In the special case of uniform $\bar{t}=u$, this is equivalent to maximizing the entropy, possibly subject to
constraints on $t$ that reflect its partial knowledge or a restricted set of feasible choices.
The application of this idea to the establishment of a privacy criterion is the object of the remainder of this work.

\subsection{Measuring the Privacy of User Profiles}
\label{sec:PrivacyCriterion:UserProfiles}

\noindent
We are finally equipped to justify, or at least interpret, our proposal to adopt Shannon's entropy and KL divergence as
measures of the privacy of a user profile.
Before we dive in, we must stress that the use of entropy as a measure of privacy, in the widest sense of the term,
is by no means new.
Shannon's work in the fifties introduced the concept of~\emph{equivocation} as the conditional entropy of
a private message given an observed cryptogram~\cite{Shannon49Bell},
later used in the formulation of the problem of the wiretap channel~\cite{Wyner75Bell,Csiszar78IT} as a measure of confidentiality.
More recent studies~\cite{Diaz02PET,Diaz05PhD} rescue
the suitable applicability of the concept of entropy as a measure of privacy,
by proposing to measure the degree of anonymity observable by an attacker as the entropy of
the probability distribution of possible senders of a given message.
More recent work has taken initial steps in
relating privacy to information\hyph theoretic quantities~\cite{Rebollo10IT,Rebollo10KDE,Rebollo08PSD,Li07ICDE}.

In the context of this paper,
an intuitive justification in favor of entropy maximization
is that it boils down to making the apparent user profile as uniform as possible,
thereby hiding a user's particular bias towards certain categories of interest.
But a much richer argumentation stems from Jaynes' rationale behind entropy maximization methods~\cite{Jaynes82P,Jaynes57PRS2},
more generally understood under
the beautiful perspective of the method of types and large deviation theory~\cite[\S 11]{Cover06B},
which we motivated and reviewed in the previous subsection.

Under Jaynes' rationale on entropy maximization methods, the entropy of an apparent user profile,
modeled by a relative frequency histogram of categorized queries,
may be regarded as a measure of privacy, or perhaps more accurately, anonymity.
The leading idea is that the method of types from information theory establishes an approximate monotonic relationship
between the likelihood of a PMF in a stochastic system and its entropy.
Loosely speaking and in our context, the higher the entropy of a profile, the more likely it is,
and the more users behave according to it.
This is of course in the absence of a probability distribution model for the PMFs, viewed abstractly as r.v.'s themselves.
Under this interpretation, entropy is a measure of anonymity, \emph{not} in the sense that the user's identity remains unknown,
but only in the sense that higher likelihood of an apparent profile, believed by an external observer to be the actual profile,
makes that profile more common, hopefully helping the user go unnoticed,
less interesting to an attacker assumed to strive to target peculiar users.

If an aggregated histogram of the population were available as a reference profile,
the extension of Jaynes' argument to relative entropy,
that is, to the KL divergence, would also give an acceptable measure of privacy (or anonymity).
Recall from Sec.~\ref{sec:Background:StatisticsAndInformationTheory} that
KL divergence is a measure of discrepancy between probability distributions,
which includes Shannon's entropy as the special case when the reference distribution is uniform.
Conceptually, a lower KL divergence hides discrepancies with respect to a reference profile, say the population's,
and there also exists a monotonic relationship between the likelihood of a distribution and
its divergence with respect to the reference distribution of choice,
which enables us to regard KL divergence as a measure of anonymity in a sense entirely analogous to the above mentioned.
In fact, KL divergence was used recently in our own work~\cite{Rebollo10IT,Parra10TB} as a generalization of entropy to measure privacy,
although the justification used built upon a number of technicalities,
and the connection to Jaynes' rationale was not nearly as detailed as in this manuscript.


\section{Application of our Privacy Criterion to Query Forgery}
\label{sec:QueryForgery}
This section applies the information\hyph theoretic privacy criterion proposed in Sec.~\ref{sec:PrivacyCriterion}
to query forgery in private information retrieval.
More specifically, Sec.~\ref{sec:QueryForgery:Criterion} establishes a privacy measure in accordance to our criterion,
which leads to the optimization problem shown in~Sec.~\ref{sec:QueryForgery:Tradeoff},
representing the compromise between privacy risk the redundancy introduced by bogus queries.
This section has been adapted from our recent work on query forgery~\cite{Rebollo10IT}, to illustrate
the criterion carefully detailed in this manuscript, and to reach a wider audience than that intended
in our original, densely mathematical work.


\subsection{Measuring the Privacy Gained by Forging Queries}
\label{sec:QueryForgery:Criterion}

\noindent
Our mathematical model represents user \emph{queries} as r.v.'s, which take on values in a common, finite alphabet.
Preliminarily, we simply model user queries as r.v.'s in a rather small set of categories, topics or keywords,
represented by $\{1,\dots,n\}$.
User profiles are modeled as the corresponding PMFs.

\begin{figure}
\centering
\includegraphics[scale=\FigScale]{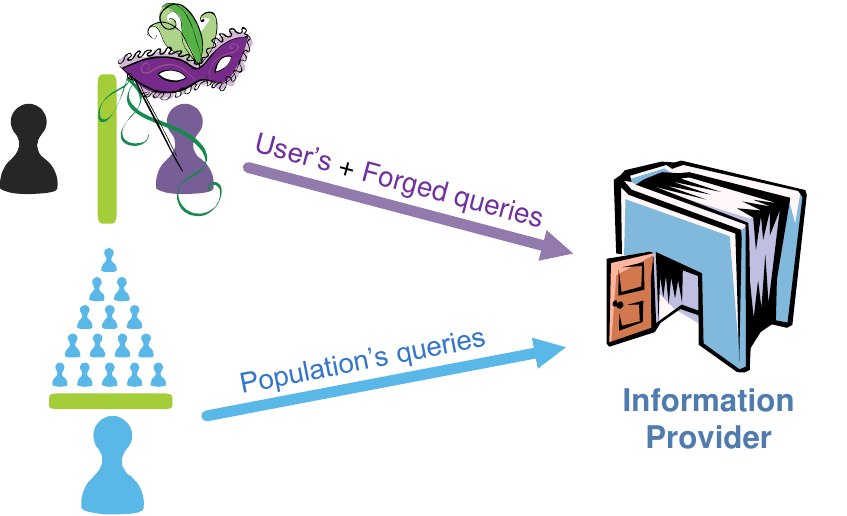}
\caption{A user accompanies original queries, submitted to an information service provider, with forged ones, in order to go unnoticed.}
\label{fig:queryforgery}
\end{figure}

Bearing in mind these considerations, we shall define~\emph{p} as the distribution of the
\emph{population}'s queries, \emph{q} as the distribution of legitimate queries of a particular \emph{user},
and \emph{r} as the distribution of queries \emph{forged} by that user.
In addition, we shall introduce a query \emph{redundancy} parameter $\rho\in[0,1)$, which will
represent the ratio of forged queries to total queries.
Concordantly,
we shall define the user's \emph{apparent} query distribution as the convex combination $s=(1-\rho)\,q+\rho\,r$,
which will actually be the distribution the information service provider, or simply a privacy attacker, will observe.
Fig.~\ref{fig:queryforgery} depicts the intuition that an attacker will be able to compromise a user's anonymity
if the user's apparent query distribution diverges too much from the population's.

Building upon the privacy criteria proposed in Sec.~\ref{sec:PrivacyCriterion}, we define the
\emph{initial privacy risk} as the KL divergence between the user's authentic profile and the population's distribution, that is,
$\oD(q\,\|\,p)$.
Similarly, we define the \emph{(final) privacy risk}~$\mathcal{R}$ as the KL divergence between the apparent distribution and
the population's, that is,
$$\cR=\oD(s\,\|\,p)=\oD((1-\rho)\,q+\rho\,r\,\|\,p).$$
We have mentioned that entropy maximization was the special case of divergence minimization when
the reference distribution is uniform.
In terms of this formulation, for a population profile $p=u$ uniform across the $n$ categories of interest,
$$\oD(s\,\|\,u)=\log n - \oH(s),$$
and, accordingly, we may regard $\oH(s)$ as a measure of privacy gain, rather than risk.


\subsection{Optimizing Privacy Subject to a Forgery Constraint}
\label{sec:QueryForgery:Tradeoff}
This section presents a formulation of the compromise between privacy and the redundant traffic due to query forgery,
which arises from the privacy measure introduced in Sec.~\ref{sec:QueryForgery:Criterion}.
Taking into account the definition of our privacy criterion, we shall suppose that the population is large enough to neglect
the impact of the choice of $r$ on~$p$.
Accordingly, we define the \emph{privacy\hyph redundancy} function
$$\cR(\rho)=\min_r \ \oD((1-\rho)\,q+\rho\,r\,\|\,p),$$
which poses the optimal trade\hyph off between query privacy (risk) and redundancy.
The minimization variable is the PMF $r$ representing the optimum profile of forged queries, for a given redundancy~$\rho$.

There are two important advantages in modeling the privacy of a user profile as a divergence in general,
or an entropy in particular, in this and other potential applications of our privacy criterion.
First, the mathematical tractability demonstrated in~\cite{Rebollo10IT}.
Secondly, the privacy\hyph redundancy function has been defined in terms of an optimization problem,
whose objective function is convex, subject to an affine constraint.
As a consequence, this problem belongs to the extensively studied class of convex optimization problems~\cite{Boyd04B} and
may be solved numerically, using a number of extremely efficient methods,
such as interior\hyph point methods.

A dual version of this problem is that of tag suppression in the semantic web~\cite{Parra10TB},
where entropy is used as a measure of privacy of user profiles, and users may choose to refrain from tagging certain
resources regarding certain categories of interest.
The privacy measure utilized may be more clearly justified, and immediately extensible to divergences, under the
considerations described in this work.

\section{Conclusion}
\label{sec:Conclusion}

\noindent
There are a wide variety of proposals for the problem of PIR, considered here in the broadest sense of the term. Within those approaches,
query forgery arises as a simple strategy in terms of infrastructure requirements, as users do not need an external entity to trust.
However, this strategy poses a trade-off between privacy and the cost of traffic and processing overhead.

In our previous work~\cite{Rebollo10IT}, we presented an
information-theoretic privacy criterion for query forgery in PIR, which arose from the formulation of the
privacy\hyph redundancy compromise. Inspired by the work in~\cite{Rebollo10KDE}, the privacy risk was measured as the KL divergence between
the user's apparent query distribution, containing dummy queries, and the population's.
Our formulation contemplated, as a special case,
the maximization of the entropy of the user's distribution.
Preliminarily, we simply model user queries as r.v.'s in a rather small set of categories, topics or keywords,
and user profiles as the corresponding PMFs.

In this work, we make a twofold contribution.
First, we thoroughly interpret and justify the privacy metric proposed in our previous work,
elaborating on the intimate connection between the celebrated method of entropy maximization
and the use of entropies and divergences as measures of privacy.
Measuring privacy enables us to optimize it, drawing upon powerful tools of convex optimization.
The entropy maximization method is a beautiful principle amply exploited in fields such as physics,
electrical engineering and even natural language processing.

Secondly, we attempt to bridge the gap between the privacy and the information\hyph theoretic
communities by substantially adapting some technicalities of our original work to reach a wider audience,
not intimately familiar with information theory and the method of types.
As neither information theory nor convex optimization are fully widespread in the privacy community,
we elaborate and clarify the connection with privacy in far more detail, and hopefully in more accessible terms, than in our original work.

Although our proposal arises from an information\hyph theoretic quantity and it is mathematically
tractable, the adequacy of our formulation relies on the appropriateness of the criteria optimized, which depends on several factors,
such as the particular application at hand, the query statistics of the users,
the actual network and processing overhead incurred by
introducing forged queries, the adversarial model and the mechanisms against privacy contemplated.



\begin{thebibliography}{10}
\providecommand{\url}[1]{#1}
\csname url@samestyle\endcsname
\providecommand{\newblock}{\relax}
\providecommand{\bibinfo}[2]{#2}
\providecommand{\BIBentrySTDinterwordspacing}{\spaceskip=0pt\relax}
\providecommand{\BIBentryALTinterwordstretchfactor}{4}
\providecommand{\BIBentryALTinterwordspacing}{\spaceskip=\fontdimen2\font plus
\BIBentryALTinterwordstretchfactor\fontdimen3\font minus
  \fontdimen4\font\relax}
\providecommand{\BIBforeignlanguage}[2]{{%
\expandafter\ifx\csname l@#1\endcsname\relax
\typeout{** WARNING: IEEEtran.bst: No hyphenation pattern has been}%
\typeout{** loaded for the language `#1'. Using the pattern for}%
\typeout{** the default language instead.}%
\else
\language=\csname l@#1\endcsname
\fi
#2}}
\providecommand{\BIBdecl}{\relax}
\BIBdecl

\bibitem{Fallows05PIALP}
D.~Fallows, ``Search engine users,'' Pew Internet and Amer. Life Project, Res.
  Rep., Jan. 2005.

\bibitem{Fox09PIALP}
S.~Fox and S.~Jones, ``The social life of health information,'' Pew Internet
  and Amer. Life Project, Res. Rep., Jun. 2009.

\bibitem{Rebollo10IT}
D.~Rebollo-Monedero and J.~Forn{\'e}, ``Optimal query forgery for private
  information retrieval,'' \emph{{IEEE} Trans. Inform. Theory}, vol.~56, no.~9,
  pp. 4631--4642, 2010.

\bibitem{Mokbel06VLDB}
M.~F. Mokbel, C.~Chow, and W.~G. Aref, ``The new {C}asper: query processing for
  location services without compromising privacy,'' in \emph{Proc. Int. Conf.
  Very Large Databases}, Seoul, Korea, 2006, pp. 763--774.

\bibitem{Duckham01CEUS}
M.~Duckham, K.~Mason, J.~Stell, and M.~Worboys, ``A formal approach to
  imperfection in geographic information,'' \emph{Elsevier Comput., Environ.,
  Urban Syst.}, vol.~25, no.~1, pp. 89--103, 2001.

\bibitem{Rebollo09IADIS}
D.~Rebollo-Monedero, J.~Forn{\'e}, L.~Subirats, A.~Solanas, and
  A.~Mart{\'i}nez-Ballest{\'e}, ``A collaborative protocol for private
  retrieval of location-based information,'' in \emph{Proc. {IADIS} Int. Conf.
  e-Society}, Barcelona, Spain, Feb. 2009.

\bibitem{Rebollo09COMCOM}
\BIBentryALTinterwordspacing
D.~Rebollo-Monedero, J.~Forn{\'e}, A.~Solanas, and T.~Martínez-Ballest{\'e},
  ``Private location-based information retrieval through user collaboration,''
  \emph{Elsevier Comput. Commun.}, vol.~33, no.~6, pp. 762--774, 2010.
  [Online]. Available: \url{http://dx.doi.org/10.1016/j.comcom.2009.11.024}
\BIBentrySTDinterwordspacing

\bibitem{Ostrovsky07PKC}
R.~Ostrovsky and W.~E. {Skeith III}, ``A survey of single-database {PIR}:
  {T}echniques and applications,'' in \emph{Proc. Int. Conf. Practice, Theory
  Public-Key Cryptogr. ({PKC})}, ser. Lecture Notes Comput. Sci. ({LNCS}), vol.
  4450.\hskip 1em plus 0.5em minus 0.4em\relax Beijing, China: Springer-Verlag,
  Sep. 2007, pp. 393--411.

\bibitem{Ghinita08MD}
G.~Ghinita, P.~Kalnis, A.~Khoshgozaran, C.~Shahabi, and K.-L. Tan, ``Private
  queries in location based services: {A}nonymizers are not necessary,'' in
  \emph{Proc. {ACM} {SIGMOD} Int. Conf. Manage. Data}, Vancouver, Canada, Jun.
  2008, pp. 121--132.

\bibitem{Elovici02WPES}
Y.~Elovici, B.~Shapira, and A.~Maschiach, ``A new privacy model for hiding
  group interests while accessing the web,'' in \emph{Proc. {ACM} Workshop on
  Privacy in the Electron. Society}.\hskip 1em plus 0.5em minus 0.4em\relax
  Washington, DC: ACM, 2002, pp. 63--70.

\bibitem{Shapira05IST}
B.~Shapira, Y.~Elovici, A.~Meshiach, and T.~Kuflik, ``{PRAW} -- {T}he model for
  {PRivAte Web},'' \emph{J. Amer. Soc. Inform. Sci., Technol.}, vol.~56, no.~2,
  pp. 159--172, 2005.

\bibitem{Frakes92PRE}
W.~B. Frakes and R.~A. Baeza-Yates, Eds., \emph{Information Retrieval: Data
  Structures {\&} Algorithms}.\hskip 1em plus 0.5em minus 0.4em\relax
  Prentice-Hall, 1992.

\bibitem{Kuflik03LNCS}
T.~Kuflik, B.~Shapira, Y.~Elovici, and A.~Maschiach, ``Privacy preservation
  improvement by learning optimal profile generation rate,'' in \emph{User
  Modeling}, ser. Lecture Notes Comput. Sci. ({LNCS}), vol. 2702.\hskip 1em
  plus 0.5em minus 0.4em\relax Springer-Verlag, 2003, pp. 168--177.

\bibitem{Howe06B}
\BIBentryALTinterwordspacing
D.~C. Howe and H.~Nissenbaum, \emph{Lessons from the Identity Trail: {P}rivacy,
  Anonymity and Identity in a Networked Society}.\hskip 1em plus 0.5em minus
  0.4em\relax NY: Oxford Univ. Press, 2006, ch. {TrackMeNot}: {R}esisting
  surveillance in web search. [Online]. Available:
  \url{http://mrl.nyu.edu/~dhowe/trackmenot}
\BIBentrySTDinterwordspacing

\bibitem{Toubiana07S}
\BIBentryALTinterwordspacing
V.~Toubiana, ``{SquiggleSR},'' 2007. [Online]. Available:
  \url{http://www.squigglesr.com}
\BIBentrySTDinterwordspacing

\bibitem{Kido05ICDE}
H.~Kido, Y.~Yanagisawa, and T.~Satoh, ``Protection of location privacy using
  dummies for location-based services,'' in \emph{Proc. {IEEE} Int. Conf. Data
  Eng. ({ICDE})}, Washington, DC, Oct. 2005, p. 1248.

\bibitem{Soghoian07ISJLP}
C.~Soghoian, ``The problem of anonymous vanity searches,'' \emph{{I/S}: J. Law,
  Policy Inform. Soc. ({ISJLP})}, Jan. 2007.

\bibitem{Samarati98SRI}
P.~Samarati and L.~Sweeney, ``Protecting privacy when disclosing information:
  $k$-{A}nonymity and its enforcement through generalization and suppression,''
  SRI Int., Tech. Rep., 1998.

\bibitem{Sun08TDP}
X.~Sun, H.~Wang, J.~Li, and T.~M. Truta, ``Enhanced $p$-sensitive $k$-anonymity
  models for privacy preserving data publishing,'' \emph{Trans. Data Privacy},
  vol.~1, no.~2, pp. 53--66, 2008.

\bibitem{Truta06PDM}
T.~M. Truta and B.~Vinay, ``Privacy protection: $p$-sensitive $k$-anonymity
  property,'' in \emph{Proc. Int. Workshop Privacy Data Manage. ({PDM})},
  Atlanta, GA, 2006, p.~94.

\bibitem{Machanavajjhala06ICDE}
A.~Machanavajjhala, J.~Gehrke, D.~Kiefer, and M.~Venkitasubramanian,
  ``$l$-{D}iversity: {P}rivacy beyond $k$-anonymity,'' in \emph{Proc. {IEEE}
  Int. Conf. Data Eng. ({ICDE})}, Atlanta, GA, Apr. 2006, p.~24.

\bibitem{JianMin08ISIP}
H.~Jian-min, C.~Ting-ting, and Y.~Hui-qun, ``An improved {V-MDAV} algorithm for
  $l$-diversity,'' in \emph{Proc. {IEEE} Int. Symp. Inform. Processing
  ({ISIP})}, Moscow, Russia, May 2008, pp. 733--739.

\bibitem{Domingo08PSAI}
J.~Domingo-Ferrer and V.~Torra, ``A critique of $k$-anonymity and some of its
  enhancements,'' in \emph{Proc. Workshop Privacy, Security, Artif. Intell.
  ({PSAI})}, Barcelona, Spain, 2008, pp. 990--993.

\bibitem{Li07ICDE}
N.~Li, T.~Li, and S.~Venkatasubramanian, ``$t$-{C}loseness: {P}rivacy beyond
  $k$-anonymity and $l$-diversity,'' in \emph{Proc. {IEEE} Int. Conf. Data Eng.
  ({ICDE})}, Istanbul, Turkey, Apr. 2007, pp. 106--115.

\bibitem{Rebollo08PSD}
D.~Rebollo-Monedero, J.~Forn{\'e}, and J.~Domingo-Ferrer, ``From $t$-closeness
  to {PRAM} and noise addition via information theory,'' in \emph{Privacy Stat.
  Databases ({PSD})}, ser. Lecture Notes Comput. Sci. ({LNCS}).\hskip 1em plus
  0.5em minus 0.4em\relax Istambul, Turkey: Springer-Verlag, Sep. 2008, pp.
  100--112.

\bibitem{Rebollo10KDE}
\BIBentryALTinterwordspacing
------, ``From $t$-closeness-like privacy to postrandomization via information
  theory,'' \emph{{IEEE} Trans. Knowl. Data Eng.}, vol.~22, no.~11, pp.
  1623--1636, Nov. 2010. [Online]. Available:
  \url{http://doi.ieeecomputersociety.org/10.1109/TKDE.2009.190}
\BIBentrySTDinterwordspacing

\bibitem{Diaz02PET}
C.~D{\'i}az, S.~Seys, J.~Claessens, and B.~Preneel, ``Towards measuring
  anonymity,'' in \emph{Proc. Workshop Privacy Enhanc. Technol. ({PET})}, ser.
  Lecture Notes Comput. Sci. ({LNCS}), vol. 2482.\hskip 1em plus 0.5em minus
  0.4em\relax Springer-Verlag, Apr. 2002.

\bibitem{Diaz05PhD}
C.~D{\'i}az, ``Anonymity and privacy in electronic services,'' Ph.D.
  dissertation, Katholieke Univ. Leuven, Dec. 2005.

\bibitem{Parra10TB}
J.~Parra-Arnau, D.~Rebollo-Monedero, and J.~Forn{\'e}, ``A privacy-preserving
  architecture for the semantic web based on tag suppression,'' in \emph{Proc.
  Int. Conf. Trust, Privacy, Security, Digit. Bus. ({TRUSTBUS})}, Bilbao,
  Spain, Aug. 2010.

\bibitem{Cover06B}
T.~M. Cover and J.~A. Thomas, \emph{Elements of Information Theory},
  2nd~ed.\hskip 1em plus 0.5em minus 0.4em\relax New York: Wiley, 2006.

\bibitem{Jaynes82P}
E.~T. Jaynes, ``On the rationale of maximum-entropy methods,'' \emph{Proc.
  {IEEE}}, vol.~70, no.~9, pp. 939--952, Sep. 1982.

\bibitem{Brillouin62B}
L.~Brillouin, \emph{Science and Information Theory}.\hskip 1em plus 0.5em minus
  0.4em\relax New York: Academic-Press, 1962.

\bibitem{Jaynes82B}
E.~T. Jaynes, \emph{Papers on Probability, Statistics and Statistical
  Physics}.\hskip 1em plus 0.5em minus 0.4em\relax Dordrecht: Reidel, 1982.

\bibitem{Burg75PhD}
J.~P. Burg, ``Maximum entropy spectral analysis,'' Ph.D. dissertation, Stanford
  Univ., 1975.

\bibitem{Berger96CL}
A.~L. Berger, J.~della Pietra, and A.~della Pietra, ``A maximum entropy
  approach to natural language processing,'' \emph{{MIT} Comput. Ling.},
  vol.~22, no.~1, pp. 39--71, Mar. 1996.

\bibitem{Shannon49Bell}
C.~E. Shannon, ``Communication theory of secrecy systems,'' Bell Syst., Tech.
  J., 1949.

\bibitem{Wyner75Bell}
A.~Wyner, ``The wiretap channel,'' Bell Syst., Tech. J.~54, 1975.

\bibitem{Csiszar78IT}
I.~Csisz{\'a}r and J.~K{\"o}rner, ``Broadcast channels with confidential
  messages,'' \emph{{IEEE} Trans. Inform. Theory}, vol.~24, pp. 339--348, May
  1978.

\bibitem{Jaynes57PRS2}
E.~T. Jaynes, ``Information theory and statistical mechanics {II},''
  \emph{Phys. Review Ser. II}, vol. 108, no.~2, pp. 171--190, 1957.

\bibitem{Boyd04B}
S.~Boyd and L.~Vandenberghe, \emph{Convex Optimization}.\hskip 1em plus 0.5em
  minus 0.4em\relax Cambridge, UK: Cambridge University Press, 2004.

\end{thebibliography}


\end{document}